# Guided Google: A Meta Search Engine and its Implementation using the Google Distributed Web Services

Ding Choon Hoong and Rajkumar Buyya

**Gri**d Computing and **D**istributed **S**ystems (GRIDS) Laboratory
Department of Computer Science and Software Engineering
The University of Melbourne, Australia
{chd, raj}@cs.mu.oz.au

## Abstract

With the advent of the Internet, search engines have begun sprouting like mushrooms after a rainfall. Only in recent years, have developers become more innovative, and came up with guided searching facilities online. The goals of these applications are to help ease and guide the searching efforts of a novice web user toward their desired objectives. A number of implementations of such services are emerging. This paper proposes a guided meta-search engine, called "Guided Google", as it serves as an interface to the actual Google.com search engine, using the Google Web Services.

## 1. Introduction

In recent years, Google has grown to be one of the most popular search engines that are available on the Web. Like most other search engines, Google indexes Web sites, Usenet news groups, news sources, etc. with the goal of producing search results that are truly relevant to the user. This is done using proprietary algorithms, which work based on the understanding that if a page is useful, other pages covering the same topic will somehow link to it. So, it can be said that Google focuses on a page's relevance and not on the number of responses. [1]

Moreover, Google allows sophisticated searches, with required and forbidden words, and the ability to restrict results based on particular language or encoding. However, only a small number of web users actually know how to utilize the true power of Google. Most average web users, make searches based on imprecise query keywords or sentences, which presents unnecessary, or worse, inaccurate results to the user. Based on this assumption, applications that help guide user's searching sessions have started to emerge. This is further motivated by the introduction of Google Web Services, which allows developers to query the Google server directly from their application.

Google has been providing access to its services via various interfaces such as the Google Toolbar and wireless searches. And now the company has made its index available to other developers through a Web services interface. This allows the developers to programmatically send a request to the Google server and get back a response. The main idea is to give the developers access to Google's functionality so that they can build applications that will help users make the best of Google. The Web service offers existing Google features such as searching, cached pages, and spelling correction. Currently the service is still in beta version and is for non-commercial use only. It is provided via SOAP (Simple Object Access Protocol) over HTTP and can be manipulated in any way that the programmer pleases. [2]

This paper proposes a meta-search engine called Guided Google that is built using the Google Web Services. Our meta-search engine guides and allows the user to view the search results with different perspectives. This is achieved through simple manipulation and automation of Google functions that are accessible from Guided Google through the Google Web Services. It provides two functionalities: to allow "Combinatorial Keyword Searching" and "Searching by Hosts". Details of how these functions work will be discussed further in section 4 of this paper.

## 2. Related Work

There has been a great deal of work done in making guided searches a reality. One of the best examples would be GuideBeam [3]. It is the result of research and development by the DSTC (Distributed Systems Technology Centre) based at the University of Queensland in Brisbane. GuideBeam works based on a principle called "rational monotonicity", which emerged from artificial intelligence research in the early nineties. In the context of GuideBeam, rational monotonicity prescribes how the user's current query can be expanded in a way which is consistent with the user's preferences for information. In other words it is a guiding principle of preferential reasoning. [4] Since users prefer certain pieces of information in their quest for information, preferential reasoning fits very nicely into the picture of guided searching. Users can intuitively navigate to the desired query in a context-sensitive manner. This is known as "Query by Navigation". The goal is to elicit a more precise query from the user, which will translate into more relevant documents being returned from the associated search engine.

Another example that is more closely related to Google would be the Google API Search Tool by Softnik Technologies [5]. It is a simple but powerful Windows software tool for searching Google. It is completely free and is not meant to be a commercial product. All one needs to do is just to register with Google for a license key and they will be entitled to conduct 1000 queries a day. Other than that, it is also an efficient research tool because it lets one easily and automatically log and record the search results. It also allows one to quickly create reports of their research. The URLs, titles etc may be copied to the clipboard and then to a spread sheet or any other software. To summarise, it lets one organize and keep track of their searching sessions, all at the convenience of their desktops.

The Google Search Tool requires the user to download and install the software before they can actually start using it. An alternative to this would be to have a web based version of that search tool. Many developers have been contributing to this effort and have made their tools freely available online. This paper itself focuses on Guided Google, which is an example of the many web based tools which will be discussed further later in this paper. Another example of this would be the Google API Proximity Search (GAPS) by Staggernation.com. It is a Perl script that uses the Google API to search Google for two search terms that appear within a certain distance from each other on a page [6]. It does this by using a seldom-discussed Google feature: within a quoted phrase, * can be used as a wildcard meaning "any word." So to search for **coppola** within 2 words of **nepotism**, in either order, 6 queries are needed[1]:

"coppola nepotism"
"coppola * nepotism"
"coppola * * nepotism"
"nepotism coppola"
"nepotism * coppola"
"nepotism * * coppola"

The GAPS script simply constructs these queries, gets the first page of Google results for each query, compiles all the results, and presents them in a specified sort order.

As can be seen, with the introduction of Google APIs, Google has paved a way for programmers to develop applications that help users better utilize the full potential of the Google Search Engine.

## 3. Architecture

The architecture of how Google Web Services interact with user applications is relatively simple (see Figure 1). The google server is still responsible for processing search queries. A programmer develops an application in any language he/she prefers (Java, C, Perl, PHP, .NET, etc.) and connects remotely to the Google Web APIs service. Communication is performed via the SOAP, which is an XML

[1] The example given is taken from Staggernation.com

(eXtensible Markup Language)-based mechanism for exchanging typed information. Once connected, the application will be able to issue search requests to Google's index of more than two billion web pages and receive results as structured data, access information in the Google cache, and check the spelling of words. Google Web APIs will support the same search syntax as the Google.com site. Figure 1 gives a brief illustration of how the applications interact with the Google server [7].

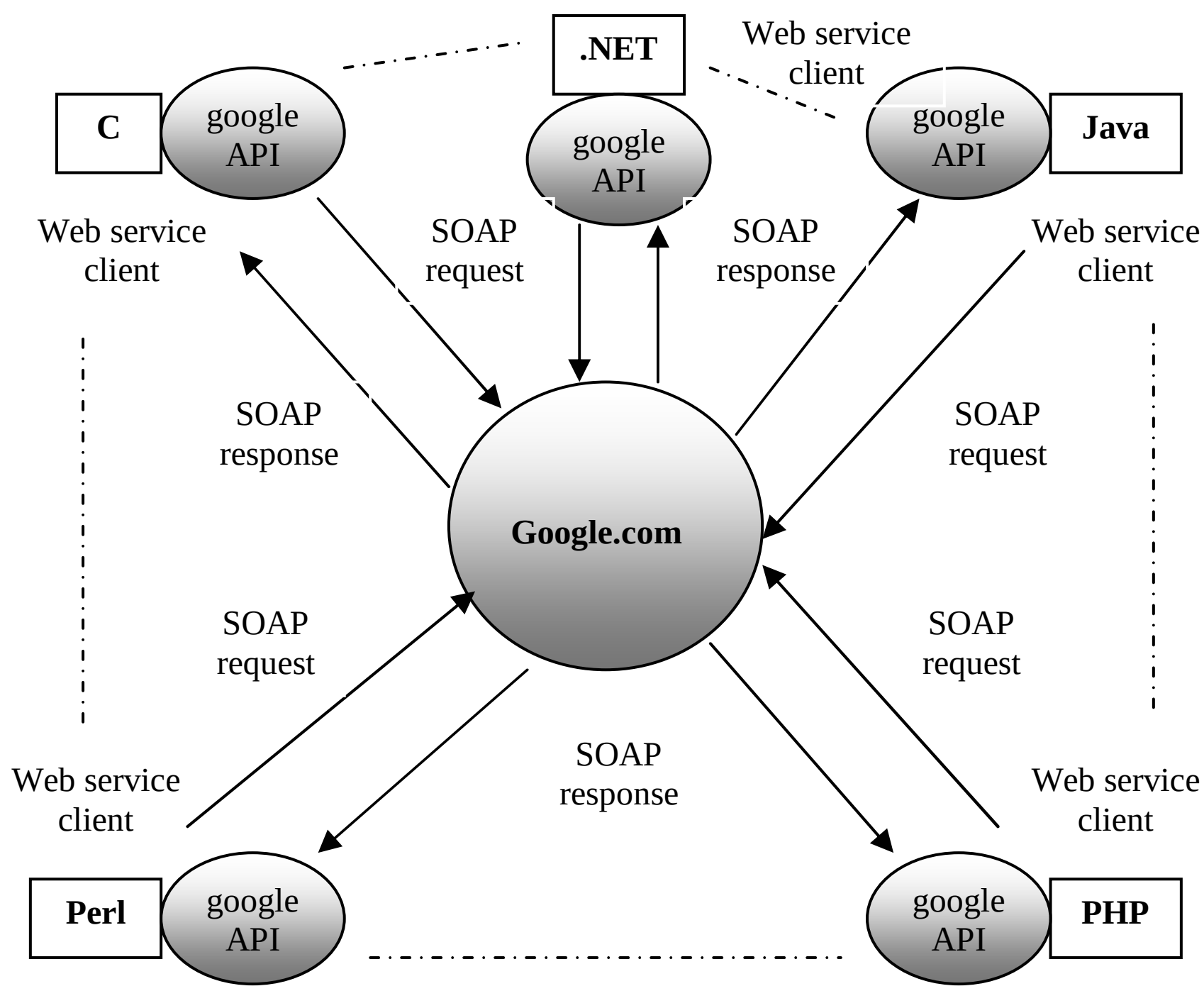

**Figure 1:** Google Web API

The interaction between the user, Guided Google and the Google server itself is illustrated in Figure 2. Basically, the user will access the Guided Google site, which was developed using JSP (JavaServer Pages) and served by a Tomcat server. Here, Tomcat functions both as a stand-alone web server and also as a servlet container. As shown before in Figure 1, Guided Google will send and receive information from the Google.com server via SOAP. This is done behind the scenes and hence, to the user, it is just like accessing a web site and sending normal HTML data.

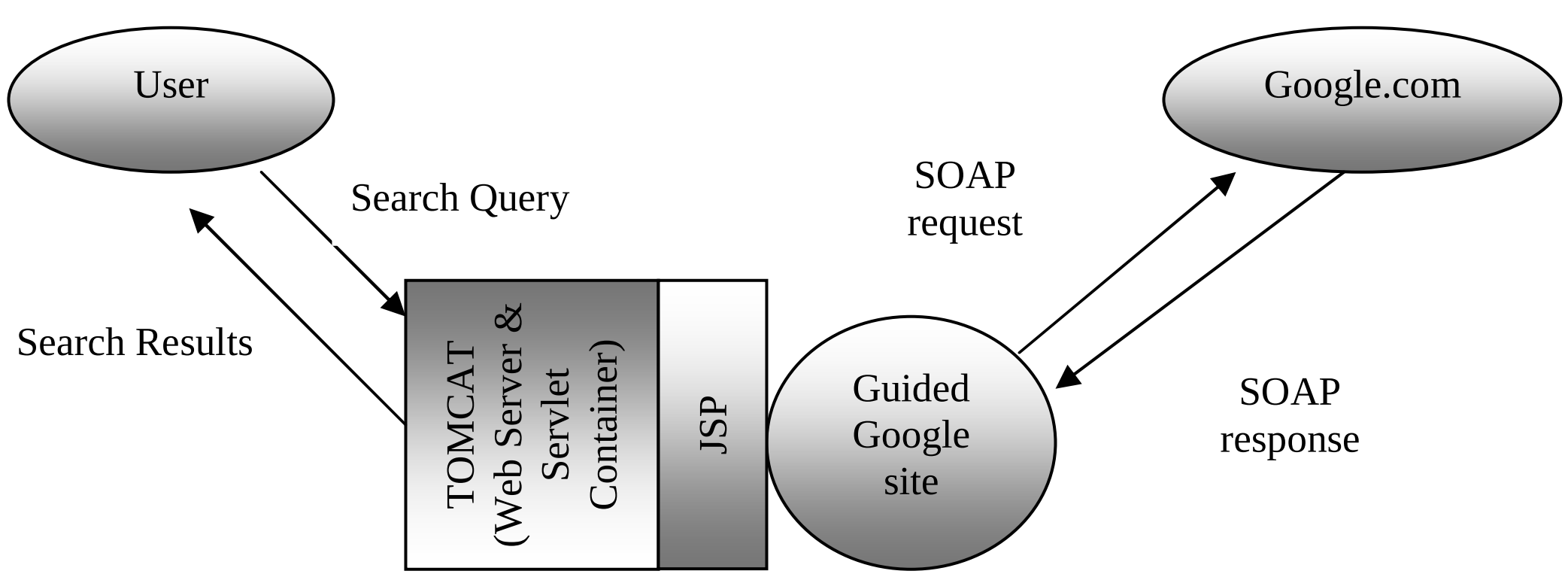

**Figure 2:** Guided Google Architecture

# 4. Design and Implementation

Guided Google is a web based guided search tool that is developed using *Forte for Java* and coded in JSP. As mentioned in the first section, this tool was developed under the assumption that an average web user makes searches based on imprecise query keywords or sentences, which in turn, presents unnecessary, or worse, inaccurate results to the user. Here, we would like to demonstrate that with just a simple tweak or manipulation of current existing Google functions, we can help guide users to get better search results.

Basically, Guided Google's functionality is split into two parts. The first one is for combinatorial keyword searching, and the second one is for keyword searching based on the servers hosting them. Each of these will be discussed in further detail later in this section. A simple illustration of how this application is structured is discussed below.

The Guided Google internal architecture is shown in the Figure 3. There is a main page (index.jsp) that is used to interface with all the other .jsp files. Normal and combinatorial keyword searching are both covered by normalSearch.jsp. Host based searching, on the other hand, is covered by two files, and they are serverBased.jsp and serverBased2.jsp. The first file merely isolates the first five unique domains from the results. serverBased2.jsp will refine the search based on the selected host and display the results in an expendable tree menu. Please refer to Figures 8 and 9 for a better illustration of how this works.

Other than that, several java bean files were also used to simplify certain tasks that needed to be performed multiple times. GoogleConnectionBean.java was used to establish the initial connection to the Google server. PermuterBean.java was used to calculate the permutation for the keywords when the user selects the combinatorial option. And ServerBean.java was used to store the URLs of unique domains which were obtained after the execution of the serverBased.jsp file. This will allow serverBased2.jsp to access those URLs when it is executed to refine the search, based on the hosts.

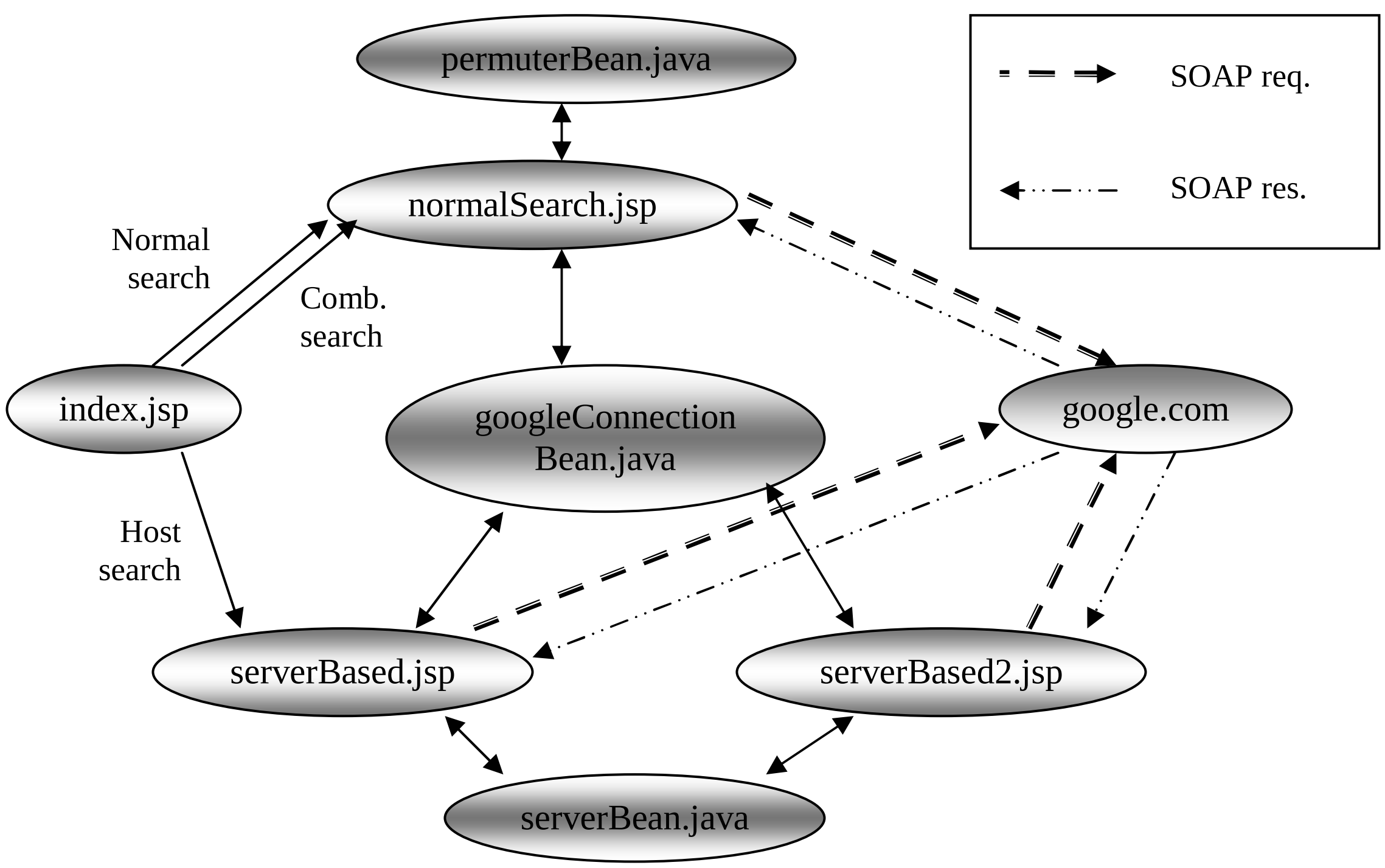

**Figure 3:** Internal Architecture of Guided Google

To simplify installation, we have exported the entire application as a WAR (Web Archive) file. In order to get it up and running, all one need to do is to have a web server that supports JSP (we are using Tomcat 3.3 here), the GoogleAPI.war file and the googleapi.jar file. We will just give brief instructions on how to setup the application [7]. However, different web servers, or even different Tomcat versions will require one to place the files differently. Instructions for setting up the environment for deploying our Guided Google are listed in Appendix A.

The remainder of this section will describe in greater detail, the two functions of Guided Google. First of all, the entry page (as illustrated in Figure 4) is made up of a query text box, a license key text box, and options as to whether the user wants to use the combinatorial or search by host functions. Users are encouraged to key in their own license key if they have registered with Google, as we are only limited to 1000 queries a day. But if they don't have one, leaving the text box 'null' would allow them to use our license key by default.

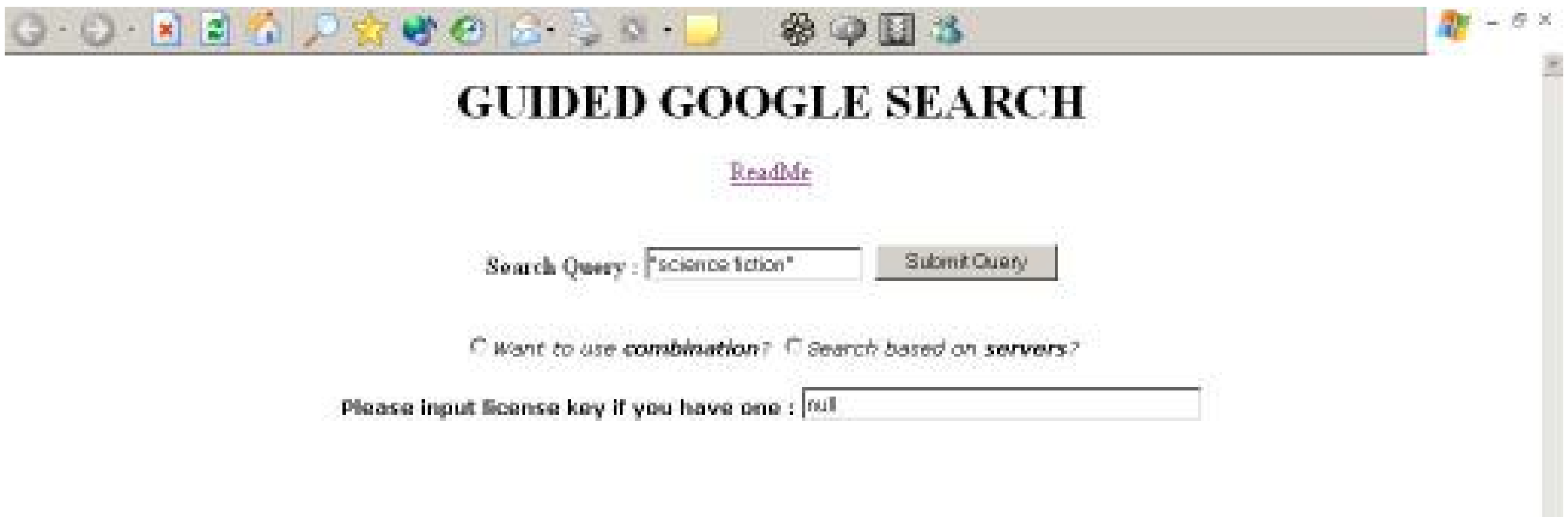

**Figure 4:** Guided Google start page

If the user submits a query keyword without selecting any of the two optional functions, Guided Google will perform a normal search (see Figure 5), as how Google.com (see Figure 6) would have done. But when one of the functions is selected, the user can get really interesting search results, as will be demonstrated in the next section. We will now explain the two Guided Google functions.

***Combinatorial Search***

It is widely known that the arrangement of keywords in the search query makes a difference in the search results of the Google Search Engine. Under the assumption that novice web users don't usually know how to construct effective keywords in their search queries; Guided Google provides a function that will automatically calculate the permutation and make different combinations of the keywords used. For example, if the user were to key in the term "science fiction", it will automatically generate two combinations; which are "science fiction" and "fiction science". The results of these two queries are very different (Please refer to Figure 8). We should also mention here that the quotes do make a difference as well. To Google; words in quotes mean that they have to occur in that particular order, in the search results. So, in accordance with this, if the search query is placed in quotes, the result of the combinations will also be reflected in quotes. The next section presents a demonstration of how this works.

***Search by Host***

This function is slightly more complex in nature. It is suppose to allow the user to search for a keyword based on the first five hosts that are hosting it. By default, Guided Google is configured to return only the first five results of a search. It is possible that there may be fewer than five hosts returned by the search. This can also happen when Guided Google automatically removes identical hosts from the list.

Beside each link is an 'arrow' sign, which means that it can be expanded. Clicking on it would make another query but this time, it is only restricted to that particular host. This can be done very easily by manipulating Google's query syntax. In this case, it would be "**site:***some_host_domain*". There are many other query syntaxes that can be very useful, depending on how the programmer manipulates them. Here, we are just showing a simple application of this in action. It is stressed here, however, that this function is not guaranteed to give more accurate results. In many cases, it won't; as the results are not the first few that were obtained from the usual keyword search query. The main purpose of this function is to allow the user to have a different perspective of searching. It gives a lateral way of looking at the results. With the search results obtained, it is hoped that the user will get a better idea of what they are searching for, and hence produce more accurate query keywords.

Other Google query syntax can also be added in. Some examples are “inurl:”, “allintext:”, “allinlinks:” etc. Each one of these can be manipulated in various ways, to give various outcomes.

## 5. Evaluation

This section is focusing on demonstrating the different search results of Guided Google. For simplicity, we will use a common search term, which is “science fiction”. This term shall be used throughout the demonstration so that the different results can be compared.

***Normal Search***

Here, “science fiction” (note that the keyword is in quotes) is submitted as the query keyword, without any options chosen, as shown in Figure 5. The results of the search are the same as a normal search done on the Google.com site. Please refer to Figure 6.

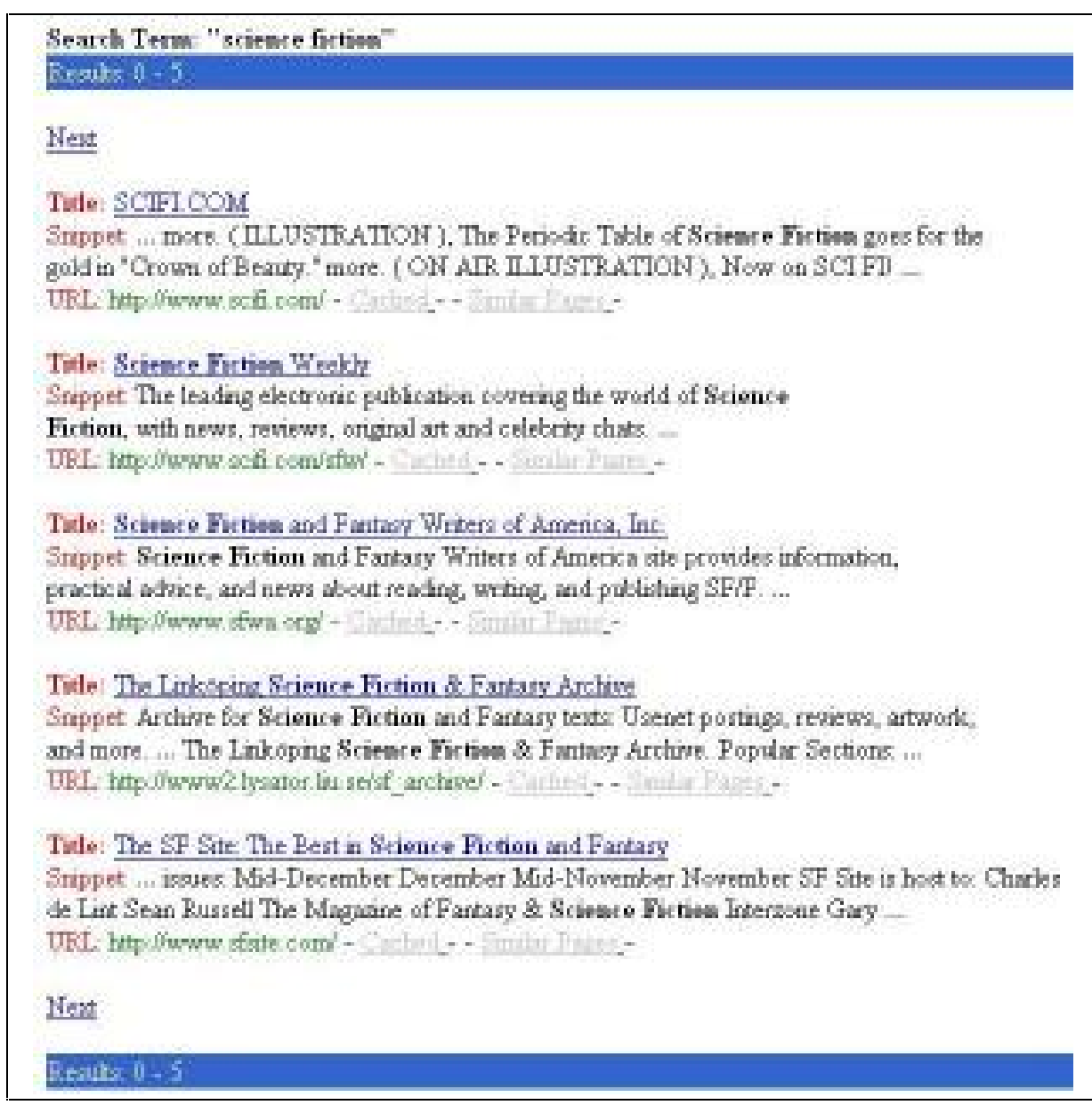

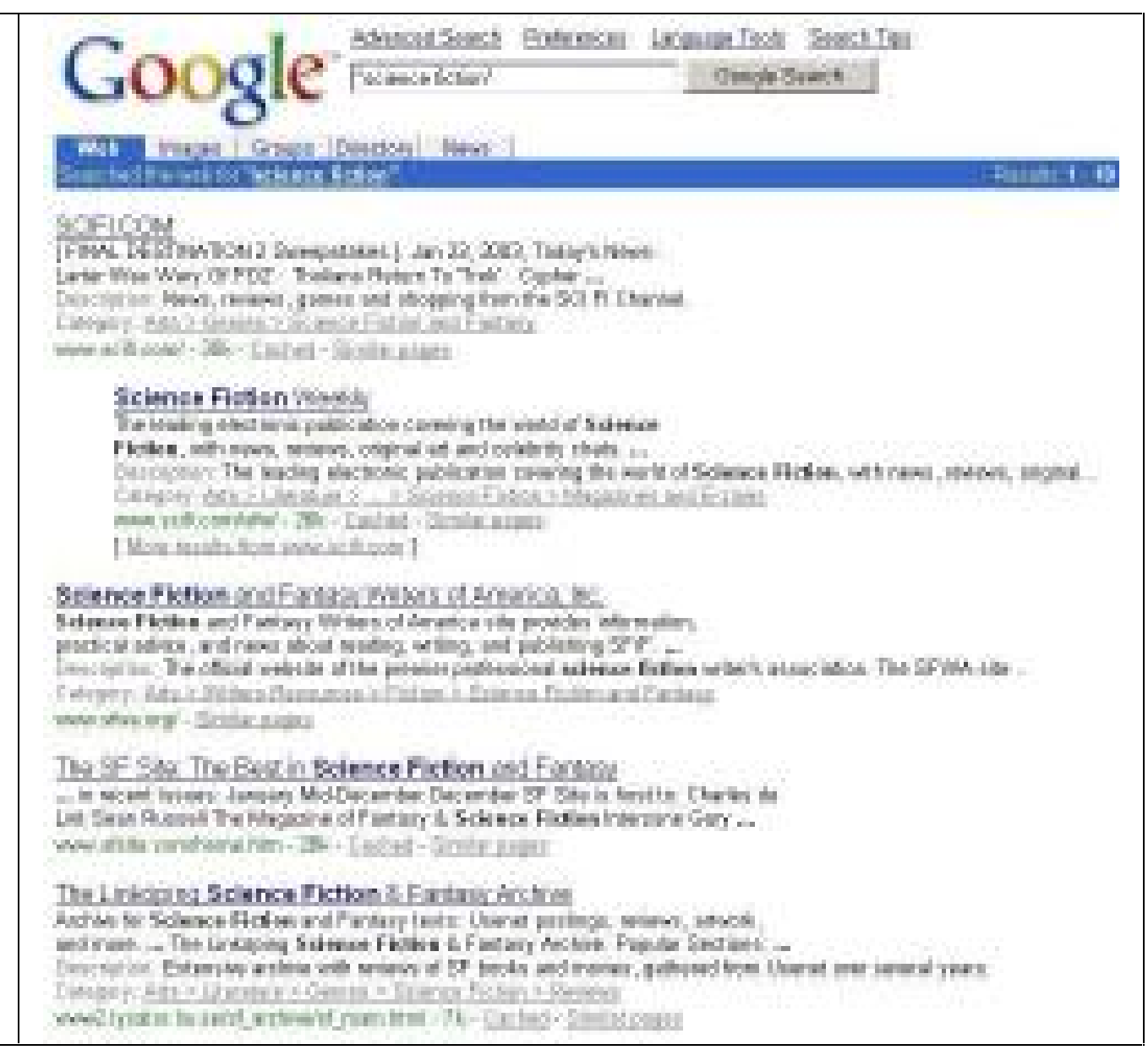

**Figure 5:** Normal Guided Google Search **Figure 6:** Normal Google Search

***Combinatorial Search***

In a combinatorial search, the term “science fiction” is passed through a permutation function which gives all unique combinations of it. In this case, it will produce “fiction science”. As permutation increases, the order of factorial, 3 keywords would mean a combination of 6 different search queries. Therefore, this function would not be practical for search queries of more than 3 words in length.

It should be noted that users are to use this function, only if they are expecting more variety of results. Words like “ski holiday” or “jet boat”, where the user is not too sure what they are looking for, may be worth searching in different sequences so that they can get a more overall result. This function automates the ‘combinations portion’ and searching for all the combinations. Figures 7 and 8 illustrate the results of two combinatorial searches for “science fiction”.

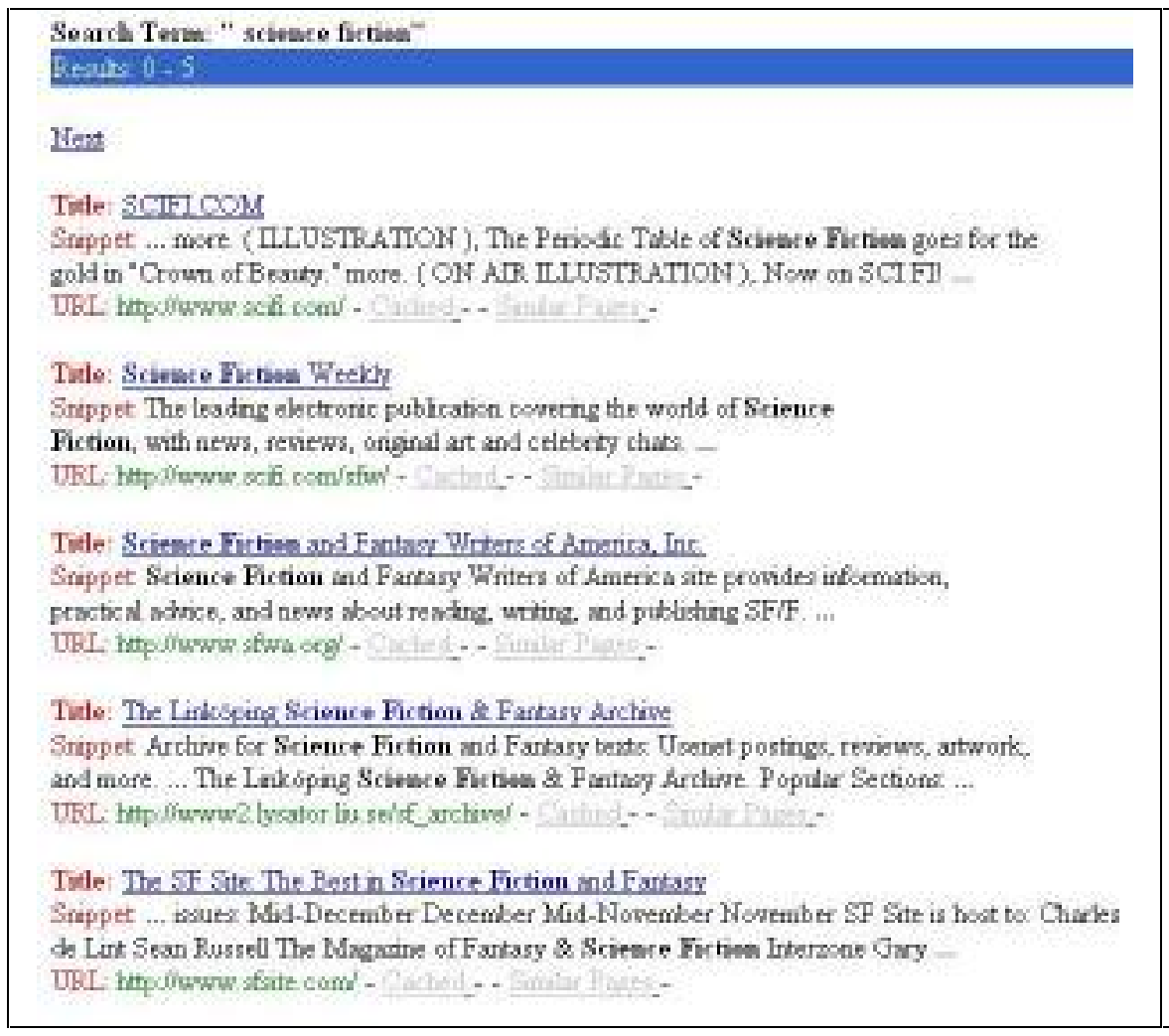

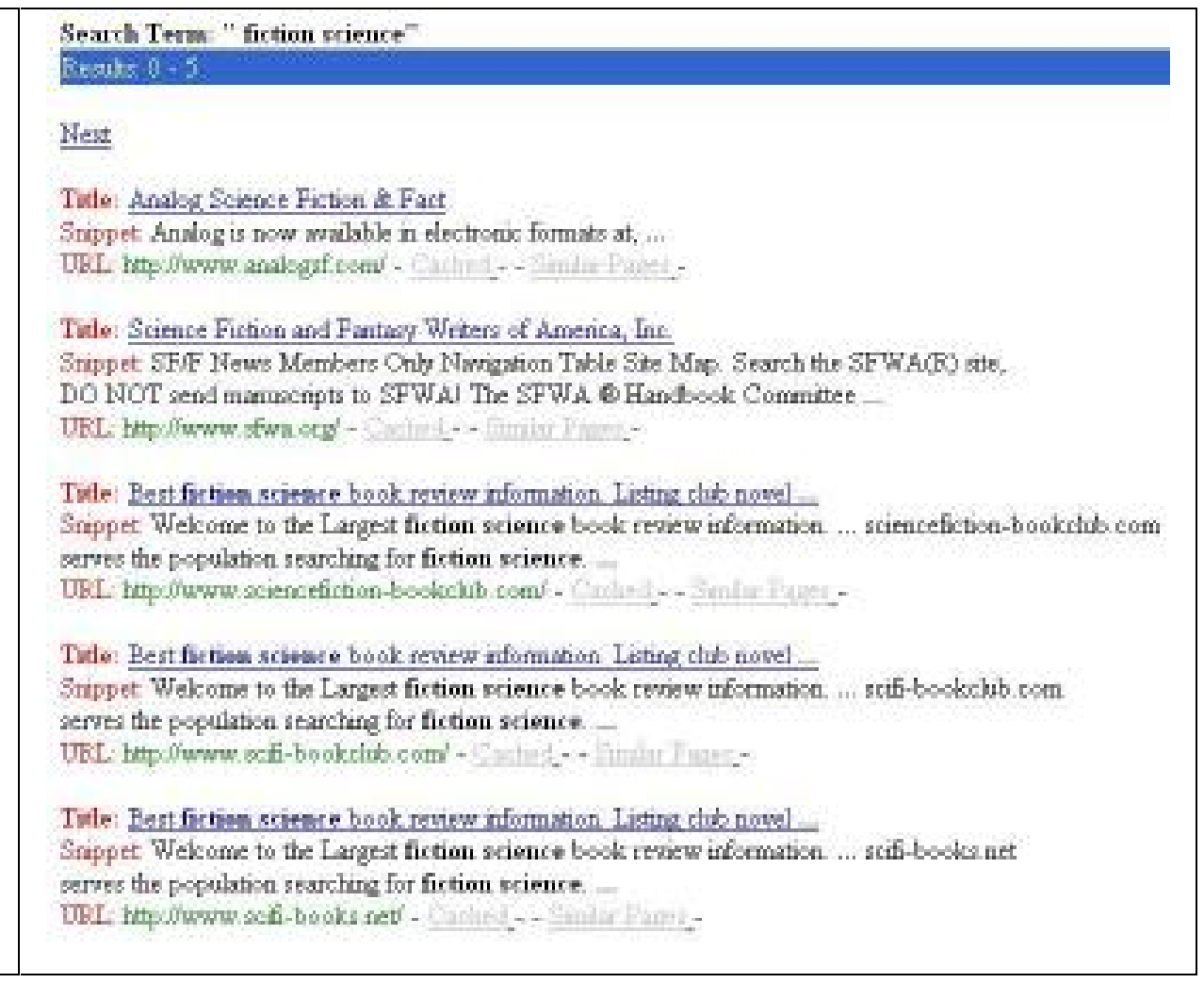

**Figure 7:** Combinatorial Search (part 1) **Figure 8:** Combinatorial Search (part 2)

## ***Search by Host***

In searching based on hosts, the term "science fiction" is submitted to Google.com and the results retrieved. Then the first 5 unique domains will be isolated and displayed to the user, as shown in Figure 9. Clicking on the arrows by the domains will submit another query to the Google site, (how this works was explained in the previous section) further refining the search based on the selected domain. Refer to Figure 10 for an illustration of this.

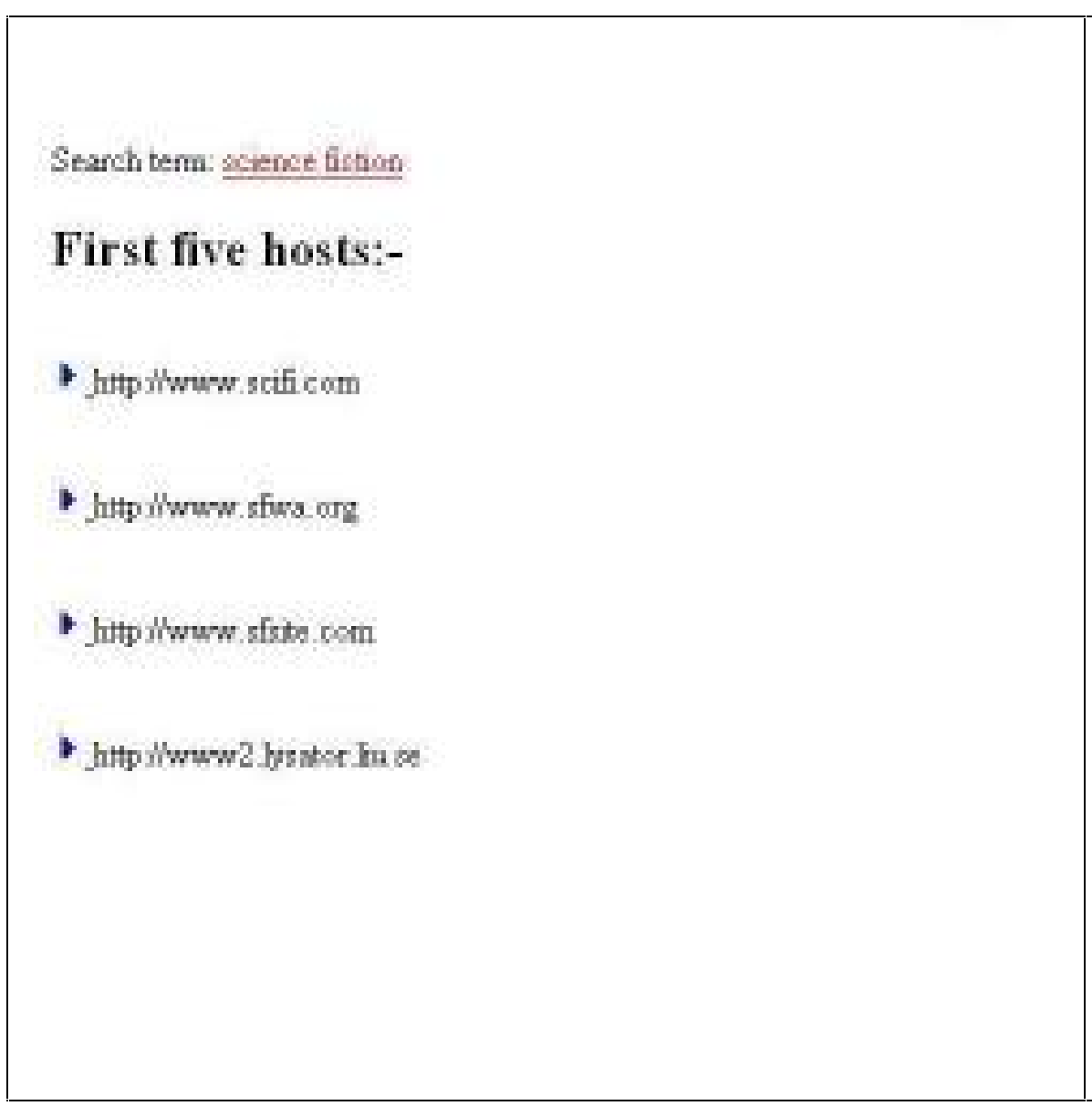

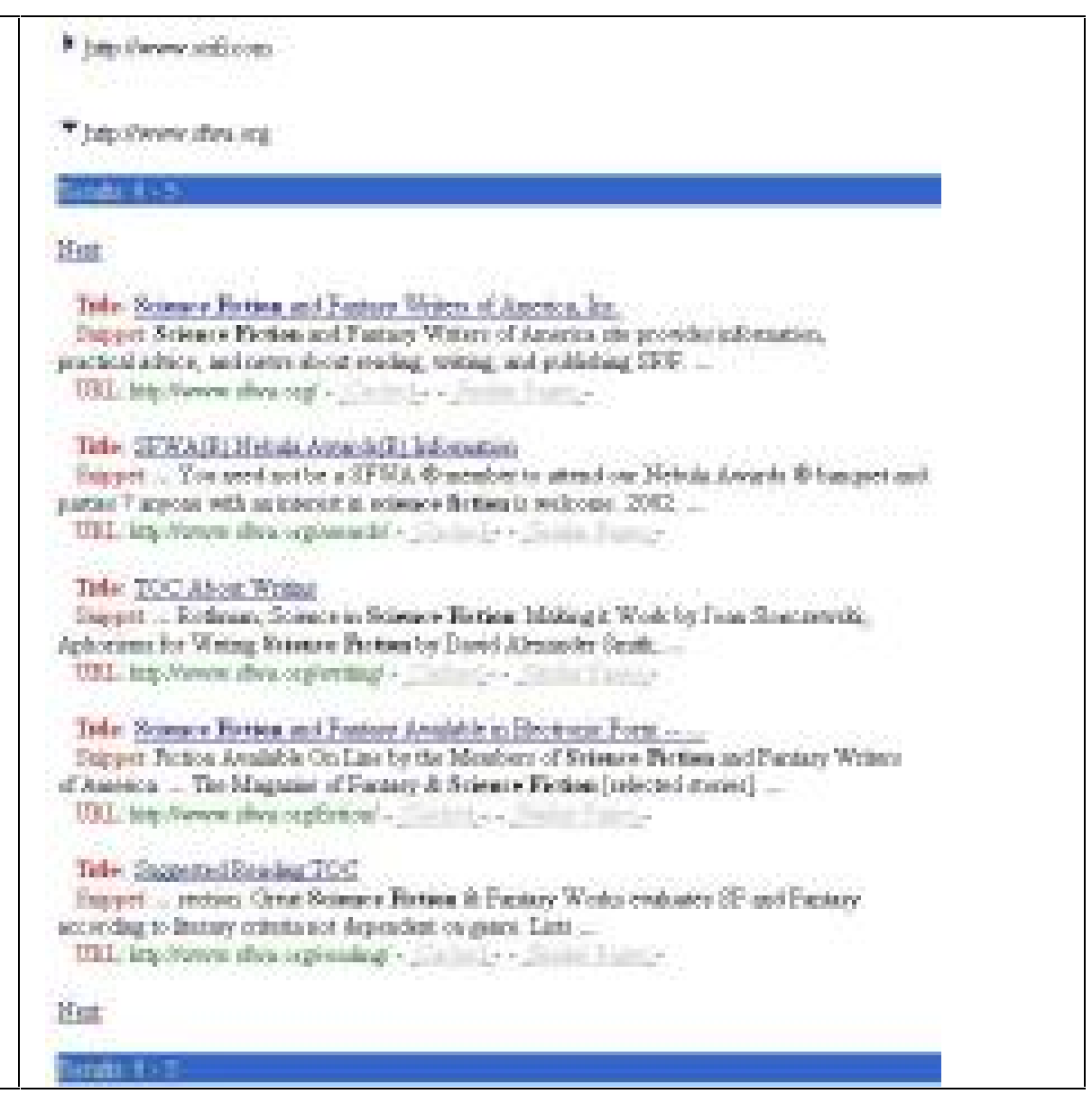

**Figure 9:** Search by Hosts (part 1) **Figure 10:** Search by Hosts (part 2)

As it can be seen, the results returned by the *'Combinatorial keyword searching'* and *'Searching by Hosts'* are very different. Even within Combinatorial searching itself, Figures 7 and 8 have already demonstrated the differences in results just by swapping the combination of search queries. Searching by Hosts; though it returns results that are not the highest in rank, this function is useful when a user wants to search for a term that is related to a certain host only.

## 6. Conclusion

This paper proposed a guided meta-search engine, called “Guided Google”, which provides meta-search capability developed using the Google Web Services. It guides and allows the user to view the search results with different perspectives. This is achieved through simple manipulation and automation of the existing Google functions. Our meta-search engine supports search based on “combinatorial keywords” and “search by hosts”. A detailed evaluation demonstrates how one can harness the capability of Google through its programmable services (Web Services).

## Acknowledgements

First and foremost, we would like to thank Google.com for making a wonderful programming interface (Google API) available that allowed us to create Guided Google. We would also like to extend our appreciation to Google Groups, for the many discussions that were held there have helped solve a few of our major problems during the implementation phase. We would like to thank Srikumar Venugopal and Anthony Sulistio for their comments on the paper.

### Appendix A: Guided Google Deployment Instructions (on Windows)

1. Download the appropriate jakarta-tomcat-<*version*> binary archive file.
2. Expand the archive into some directory (say C:\). This should create a new subdirectory named "jakarta-tomcat-<*version*>".
3. Set the environment variable JAVA_HOME to point to the root directory of your JDK hierarchy.
   set JAVA_HOME=c:/jdk1.3.1
   set PATH=%JAVA_HOME%\bin;%PATH%
4. Set the TOMCAT_HOME environment variable.
   set TOMCAT_HOME=c:\jakarta-tomcat-<*version*>
5. Place the GoogleAPI.war file into the webapps directory (c:\jakarta-tomcat-*3.3.1/webapps)*. When Tomcat is started, it will automatically expand and deploy that file.
6. Place the googleapi.jar file into the tomcat library folder. (c:\jakarta-tomcat-3.3.1\lib\apps)
7. Start Tomcat. This is done by running the bin\startup.bat file in the bin directory of Tomcat.
8. To load the main page, type http://localhost:8080/GoogleAPI/index.jsp into the browser window.
9. To stop Tomcat, just run the bin\shutdown.bat file.